\documentclass[12pt]{article}

\usepackage{cite}
\usepackage{amsmath,amsfonts}
\usepackage{amssymb}
\usepackage[usenames,dvips]{color}
\usepackage{mymacros}
\usepackage{bbm}
\usepackage{axodraw2}
\usepackage{hyperref}
\usepackage{fancyvrb}
\usepackage{enumitem}
\usepackage{caption}
\usepackage{subcaption}

\usepackage{slashed}

\usepackage{fullpage}

\renewcommand{\vec}[1]{\boldsymbol{#1}}

\newcommand{\A}{\mathcal{A}}

\makeatletter
\@addtoreset{equation}{section}
\makeatother

\makeatletter
\newcommand\footnoteref[1]{\protected@xdef\@thefnmark{\ref{#1}}\@footnotemark}
\makeatother

\newcommand{\ben}{\begin{enumerate}}
\newcommand{\een}{\end{enumerate}}

\newcommand\tra{^{\mathpalette\raiseT\intercal}}

\newcommand\raiseT[2]{%
\setbox0\hbox{$#1{#2}$}\raise\dp0\box0}

\begin{document}
\title{Feynman integral reduction by covariant differentiation}
\author{Gero von Gersdorff and Vinícius Lessa \\
\normalsize Pontif\'icia Universidade Cat\'olica, Rio de Janeiro, Brazil}
\date{}
\maketitle

\begin{abstract}
We show how a large class of Feynman integrals can be efficiently reduced to master integrals by suitable covariant differentiation on the vector space dual to the one spanned by the master integrals.  
The connections needed in the covariant derivatives have to be built only once for a given topology and then apply to any configuration of internal propagator masses.
We implement our algorithm in the {\tt Mathematica} code Method for Reduction of Loop Integrals ({\tt  MERLIN}).
\end{abstract}

\tableofcontents

\section{Introduction}

Any  momentum integral arising in perturbation theory of relativistic quantum field theories  can be written as a linear combination of so-called master integrals \cite{Chetyrkin:1981qh,Laporta:2000dsw}, 
with coefficients that are rational functions of Lorentz-invariants such as masses and scalar products of external momenta. 
The number of master integrals is finite \cite{Smirnov:2010hn,Lee:2013hzt,Bitoun:2017nre}, and they can be computed systematically using the method of differential equations \cite{Kotikov:1990kg,Remiddi:1997ny,Henn:2013pwa}.
The main task is then to find the above mentioned coefficients, which is usually done by a well-established but relatively complex algorithm \cite{Laporta:2000dsw} that has been implemented in various computer codes \cite{Anastasiou:2004vj,Studerus:2009ye,vonManteuffel:2012np,Lee:2012cn,Lee:2013mka,Smirnov:2008iw,Smirnov:2013dia,Smirnov:2014hma,Smirnov:2019qkx,Smirnov:2025prc,Maierhofer:2017gsa,Maierhofer:2018gpa,Klappert:2020nbg}. Even for quite simple cases, this algorithm needs significant computing time, and it is therefore interesting to explore alternatives \cite{Kosower:2018obg,Feng:2025leo,Song:2025pwy,Zeng:2025xbh,Liu:2025udl,delaCruz:2026mas,Smith:2025xes,Shih:2026jfe}.

The aim of the present paper is to develop a novel way of reducing Feynman integrals to linear combinations of master integrals. It is based on a differential operator (or covariant derivative) $d+A$, similar to the one that is needed to calculate the master integrals themselves by the method of differential equations. Even though this differential operator still needs to be found by the standard algorithm, this step can be done once and for all for any given topology of diagrams. We implement our method in  the {\tt Mathematica} code {\tt MERLIN} that in this version provides an initial library of differential operator for some simple topologies of up to three loops. 

An important subclass of Feynman integrals are so-called vacuum diagrams (also known as vacuum bubbles). These are diagrams that do not possess any external momenta. 
Their importance stems from their appearance in the  "running and matching" procedure of effective field theories (EFTs) \cite{Georgi:1993mps,Smirnov:2002pj,Manohar:2018aog}.
Due to the fact that EFTs amount to a local derivative expansion, one  needs the expansion of diagrams around vanishing momenta. The coefficients of this expansion  precisely correspond to vacuum bubbles. 
Our method is particularly well suited for these vacuum diagrams.

This paper is organized as follows. In section \ref{sec:covdiff} we prove our general method, which is illustrated in section \ref{sec:ex} with some simple examples.
We introduce our code \texttt{MERLIN} in section \ref{sec:merlin} and present some conclusions and outlook for future work in section \ref{sec:conclusions}.

\section{Reduction by covariant differentiation}
\label{sec:covdiff}

\begin{figure}
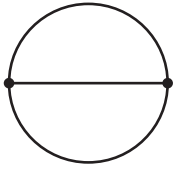

	\centering
	\scalebox{2}{
	\begin{axopicture}(30,30)(0,0)	
	\SetWidth{0.5}
    \SetColor{Black}
	\Arc(15,15)(15,0,360)
	\Line(0,15)(30,15)
	\Vertex(0,15){1}
	\Vertex(30,15){1}
\end{axopicture}
}
\caption{Two loop vacuum diagram.}
\label{fig:sunset}
\end{figure}

Consider a set of master integrals which we group into a vector $\vec I$ of dimension $N$. They are functions of the  squared masses $u_i$, $i=1\dots n$ of the $n$ internal propagators. For the time being we take the $u_i$ generic, i.e., all different.
If we were after this most general mass configuration we could easily obtain all the integrals $K$ with higher powers of propagators by simple differentiation with respect to the masses of a suitable master integral $I_a$:
\be
K(u_i)=\prod_i \frac{(-\partial_i)^{n_i}}{n_i!} I_a(u_i)\,.
\label{eq:partial}
\ee
However, the calculation of the master integrals $\vec I$   is a formidable task, and the resulting multi-scale functions are very complicated.
Most often we are however interested in a non-generic mass configuration, sometimes with only one or two mass scales. 
For instance, one might be interested in the mass configuration $(u_1,u_2,u_3)=(u,u,w)$ for the case of the two-loop sunset vacuum graph in fig.~\ref{fig:sunset}. 
Let us call this†  configuration of interest $u_{0,i}$ and define
\be
\vec I_0\equiv \vec I(u_{i,0})\,.
\ee
Due to the nongeneric mass configuration, a given integral may possess enhanced  symmetries. For instance, for the configuration $(u,u,w)$ for the two-loop sunset vacuum graph,  
the integral is symmetric under the exchange of the last two propagators.
Due to these symmetry relations the vector $\vec I_0$ contains fewer than $N$ independent entries. 
We will   denote the vector of {\em independent} master integrals $\vec J$ (with dimension $M\leq N$), and introduce an $N\times M$ matrix  $Q_0$  such that 
\be
\vec I_0=Q_0\vec J\,.
\label{eq:QJ}
\ee
The $\vec J$ are precisely the 
master integrals one needs to compute for the mass configuration of interest.
 This  set  is not only smaller, but the integrals themselves are easier to calculate and considerably simpler functions of the masses.
It would therefore be quite advantageous if one could compute integrals with higher propagator powers  by  differentiation of the reduced master integrals $\vec I_0$ or $\vec J$, circumventing the complicated reduction by IBP identities. 
Of course a naive differentiation would not work, as such a derivative now acts on several propagators at once.
The purpose of this paper is to show how this can nevertheless be done.

The generic set of master integrals (still with all masses different for now) satisfy a differential equation
\be
\partial_i\vec I=-A_i\vec I\,,
\ee
where the $A_i$ are $N\times N$ matrices whose entries are rational functions of the $u_i$ and the spacetime dimension $d$.\footnote{To find the connections $A_i$, all one needs to do is to reduce $\partial_i I_a$ ($a=1\dots N$) again to a linear combination of master integrals.}
 For the case of vacuum diagrams this is just the usual differential system used to  find the actual master integrals  \cite{Kotikov:1990kg,Remiddi:1997ny,Henn:2013pwa}, for non-vacuum diagrams  it is a subsystem where only the internal masses $u_i$ are kept as variables.
We write this as
\be
D_i\vec I=0\,,
\label{eq:diffcov}
\ee
where we defined the "covariant derivative"
\be
D_i\vec I\equiv(\partial_i+A_i)\vec I\,.
\ee
The covariant derivative satisfy the integrability conditions $[D_i,D_j]=0$ \cite{Abreu:2022mfk}. 
The matrices $A_i$ can be calculated easily with the help of publicly available codes \cite{Anastasiou:2004vj,Studerus:2009ye,vonManteuffel:2012np,Lee:2012cn,Lee:2013mka,Smirnov:2008iw,Smirnov:2013dia,Smirnov:2014hma,Smirnov:2019qkx,Smirnov:2025prc,Maierhofer:2017gsa,Maierhofer:2018gpa,Klappert:2020nbg}, and this step  has to be done only once for any given graph topology.

Let $\{\vec e_a\}$ be a basis of unit vectors of the dual vector space (which can be naturally understood as the space of coefficients of the master integrals), then each master integral can be trivially written as
\be
I_a=\vec e_a\cdot \vec I\,,
\ee 
where $\vec e_a$ is the unit vector in the direction $a$. 
For a unique choice of $n_i$ and $\vec e_a$,  any integral $K$ with higher powers of propagators than the ones present in $\vec I$ can be written as 
\be
K(u_i)=\prod_i \frac{(-\partial_i)^{n_i}}{n_i!} I_a(u_i)=\left(\prod_i \frac{(-D_i)^{n_i}}{n_i!}\vec e_a\right)\cdot \vec I(u_i)
\label{eq:covI}
\ee
where we used eq.(\ref{eq:diffcov}) and defined the contravariant derivative on the dual space
\be
D\vec e_a=(\partial-A\tra)\vec e_a
\ee
We will not distinguish in notation between the covariant and contravariant derivative, it will always be clear from the context.

Therefore, the vector 
\be
\vec X(u_i)\equiv \left(\prod_i \frac{(-D_i)^{n_i}}{n_i!}\vec e_a\right)
\label{eq:X}
\ee
provides the coefficients for the integral $K$ in terms of the master integrals $\vec I$.

For now, eq.~(\ref{eq:covI}) appears to be just a fancy rewriting of 
 eq.~(\ref{eq:partial}). Note however that the covariant derivatives no longer act on the master integral $I_a$ themselves but rather only on the connection $A_i$ which are relatively simple rational functions. 
Naively, we could then just evaluate $\vec X$ and $\vec I$ at the non-generic mass configuration, in particular, the components of $\vec I$ will be written in terms of the reduced set of simpler master integrals.
However, the limit
\be
K(u_{0,i})=\lim_{u_i\to u_{0,i}}\vec X(u_i)\cdot \vec I(u_i)
\ee
 turns out to be a little bit subtle. 
Even  though the function $K(u_i)$ (and in particular the master integrals $\vec I(u_i)$) are regular at $u_0$,  the limit $\lim_{u_i\to u_{0,i}}X(u_i)$ is in general  singular.
 
In order to take the limit properly, let us chose a direction $v$ in the mass-space, and write
 \be
 u_i=u_{0,i}+v_i t
 \label{eq:limitdir}
 \ee
 where $t$ is some parameter that we want to take to zero.
Then
\be
K(u_{0,i})=\lim_{t\to 0}\left[\vec X(t)\cdot \vec I(t)\right]
\label{eq:limit}
\ee
is the integral we are after. In general, this limit exists and is unique (i.e.), independent of the direction $v$.\footnote{A caveat concerns infrared divergences, i.e., when some of the $u_i^0$ are zero. In this case we replace the zero masses with a common regulator mass such that all IR divergences are contained  within the master integrals themselves.}

To take the limit, we may expand both $\vec X$ and $\vec I$ in $t$. As commented above, $\vec X(t)$ is in general singular, i.e. 
\be
\vec X(t)=t^{-k}\vec X_{-k}+t^{-k+1}\vec X_{-k+1}+\dots
\label{eq:X-expansion}
\ee
 for some positive integer $k$. This expansion can be straightforwardly done in any specific case.
Next we  need to expand $\vec I(t)$ up to order $t^k$  
which can be performed  as follows.\footnote{See refs.~\cite{Lee:2017qql,Armadillo:2022ugh,Lee:2018ojn} for related ideas.} First, define
\be
\A(t)\equiv v^i A_i(t)
\ee
We can use the equation
\be
\left(\frac{d}{dt}+\A(t)\right)\vec I(t)=0
\label{eq:master}
\ee
in order to arrive at a recursive formula.
The master integrals have a regular behaviour as $t\to0$, as the resulting integrals are simply the master integrals with some relations between the masses. We therefore make the ansatz 
\be
{\vec I}(t) = \vec I_{0} + t \vec I_1+ t^2 \vec I_2+\dots 
\label{eq:Iexp}
\ee
Furthermore, we will assume that the expansion of the matrix $\A$ takes the form\footnote{Eqns.~(\ref{eq:master}) and (\ref{eq:Iexp}) imply that 
$\A$ must have an expansion $\A=\sum_{n=\ell}^\infty \A_{n}t^n $ for some  integer $\ell$. In all the examples that we considered, $\ell\geq-1$, which we will assume in the following for simplicity. If $\ell<-1$ the formalism could easily be generalized.}
\be
\A(t)= t^{-1} \A_{-1}+\A_0+t\A_1 +\dots
\label{eq:Aexp}
\ee
we straightforwardly get the relation (comparing coefficients of $t^{n-1}$)
\be
\vec I_n=
-\frac{1}{n+\A_{-1}}\left[
\A_{0}\vec I_{n-1}+\dots +\A_{n-1}\vec I_0
\right]
\qquad n>0
\label{eq:recursion}
\ee
In particular, solving for the recursion,
\begin{align}
\vec I_1={}&-\frac{1}{1+\A_{-1}}\A_0\vec I_0\,,\label{eq:I1}\\
\vec I_2={}&\frac{1}{2+\A_{-1}}\left[\A_0\frac{1}{1+\A_{-1}}\A_0-\A_1\right]\vec I_0\,,\label{eq:I2}
\end{align}
etc.
Note that the expansion coefficients are expressed entirely in terms of the reduced set of master integrals $\vec I_0$. 
Moreover, making use of the relation
eq.~(\ref{eq:QJ}) 
we write
\be
\vec I(t)=Q(t)\vec J
\ee
where $Q(t)=\sum_n t^n Q_n$, and the $Q_n$ satisfy the same recursion relation eq.~(\ref{eq:recursion}) with  $\vec I_n\to Q_n$, explicitly
\begin{align}
Q_1={}&-\frac{1}{1+\A_{-1}}\A_0Q_0\,,\label{eq:I1}\\
Q_2={}&\frac{1}{2+\A_{-1}}\left[\A_0\frac{1}{1+\A_{-1}}\A_0-\A_1\right]Q_0\,,\label{eq:I2}
\end{align}
etc.

With the expansion coefficients  $Q_n$ and $\vec X_{n}$ at hand, we can easily obtain the limit eq.~(\ref{eq:limit}),
\be
K=\left(\sum_{n=0}^k \vec X\tra_{-n}\cdot Q_n\right)\cdot \vec J 
\ee

One immediate consequence of  the expansions eq.~(\ref{eq:Iexp}) and eq.~(\ref{eq:Aexp}) inserted in eq.~(\ref{eq:master}) (from the term of order $t^{-1}$) is that 
\be
A_{-1}\vec I_0=0
\label{eq:An1I0}
\ee
The  symmetries of the reduced set of master integrals $\vec I_0$ must  enforce this equation.
However, sometimes  trivial symmetries of the diagrams  can be combined with IBP identities to yield new nontrivial symmetry relations.
Eq.~(\ref{eq:An1I0}) can then be used  to find these implicit symmetries.
An example of this will be provided by the three-loop graph of section \ref{sec:example3loop}.

\section{Examples}
\label{sec:ex}

\subsection{Two-loop vacuum diagrams}
\label{sec:twoloopex}

Before introducing our code, 
let us illustrate the method by some examples.
Consider first the two-loop vacuum diagrams
\be
I_{n_1n_2n_3}\equiv \int\frac{d^dq_1}{(2\pi)^d}\frac{d^dq_2}{(2\pi)^d}\prod_{i=1}^3\frac{1}{(k_i^2+u_i)^{n_i}}
\ee
where $k_1=q_1$, $k_2=q_2$, and $k_3=q_1+q_2$. These integrals contain both possible topologies, sunset (for all $n_i>0$) and figure-eight (exactly one of the $n_i=0$ and the other positive).
There are 4 master integrals, that may be taken as 
\be
\vec I=\begin{pmatrix}
I_{011},I_{101},I_{110},I_{111}
\end{pmatrix}\tra
\ee
Taking for example the mass configuration $u_1=u_2=u$, $u_3=w$ the reduced set of master integrals can be taken as 
\be
\vec J=(J_{011},J_{110},J_{111})\tra
\ee
where $J_{abc}(u,w)=I_{abc}(u,u,w)$.
Taking the direction $v=(1,0,0)$, we get
\be
\A(t)=\begin{pmatrix}
0&0&0&0\\
0&\frac{2-d}{2(u+t)}&0&0\\
0&0&\frac{2-d}{2(u+t)}&0\\
\frac{d-2}{q}&\frac{-(d-2)(2u-w+t)}{2(u+t)q}&\frac{-(d-2)(w+t)}{2(u+t)q}&\frac{-(d-3)(-w+t)}{q}
\end{pmatrix}
\ee
where 
$q=w(w-4u) -wt+t^2$.
Notice that in this particular case $\A_{-1}=0$, such that the condition eq.~(\ref{eq:An1I0}) is automatically fulfilled and we do not get any implicit symmetries.

Using eqns.~(\ref{eq:I1}), (\ref{eq:I2}) one gets the immediate expansion
\begin{align}
Q(t)={}&
\left(
\begin{array}{ccc}
 1 & 0 & 0 \\
 1 & 0 & 0 \\
 0 & 1 & 0 \\
 0 & 0 & 1 \\
\end{array}
\right)+t
\left(
\begin{array}{ccc}
 0 & 0 & 0 \\
 \frac{d-2}{2 u} & 0 & 0 \\
 0 & \frac{d-2}{2 u} & 0 \\
 \frac{(d-2) (2 u-w)}{2 u \left(w^2-4 u w\right)}-\frac{d-2}{w^2-4 u w} & -\frac{d-2}{2 u (4 u-w)} & -\frac{d-3}{w-4 u} \\
\end{array}
\right)
+\dots
\end{align}
However, for the mass configuration $(u,u,w)$, neither $\A(t)$ nor $\vec X(t)$ are ever singular, so one really only needs the leading term $Q_0$ of this expansion.
A mass configuration that does give a singular behavior would for instance be provided by $(u,u,4u)$. We will not spell out this case here.

\subsection{Three-loop vacuum diagrams}
\label{sec:example3loop}

\begin{figure}
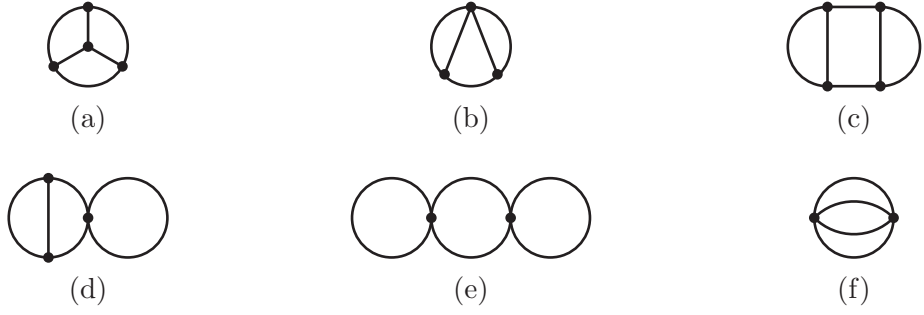

\begin{center}

\begin{subfigure}[b]{0.3\textwidth}
	\centering
	\begin{axopicture}(30,40)(0,0)	
	\SetWidth{1.0}
    \SetColor{Black}
	\Arc(15,15)(15,0,360)
	\Line(15,30)(15,15)
	\Line(15,15)(2,7.5)
	\Line(15,15)(28,7.5)
	\Vertex(15,15){2}
	\Vertex(15,30){2}
	\Vertex(2,7.5){2}
	\Vertex(28,7.5){2}		
\end{axopicture}
\caption{}
\label{fig:tetra}
\end{subfigure}
\begin{subfigure}[b]{0.3\textwidth}
	\centering
	\begin{axopicture}(30,30)(0,0)	
	\SetWidth{1.0}
    \SetColor{Black}
	\Arc(15,15)(15,0,360)
	\Line(15,30)(5,4.5)
	\Line(15,30)(25,4.5)
	\Vertex(15,30){2}
	\Vertex(5,4.5){2}
	\Vertex(25,4.5){2}		
\end{axopicture}
\caption{}
\label{fig:last}
\end{subfigure}
\begin{subfigure}[b]{0.3\textwidth}
	\centering
	\begin{axopicture}(50,30)(0,0)	
	\SetWidth{1.0}
    \SetColor{Black}
	\Arc(15,15)(15,90,270)
	\Arc(35,15)(15,270,90)
	\Line(15,0)(15,30)
	\Line(35,0)(35,30)
	\Line(15,0)(35,0)
	\Line(15,30)(35,30)
	\Vertex(15,0){2}
	\Vertex(15,30){2}
	\Vertex(35,0){2}
	\Vertex(35,30){2}
\end{axopicture}
\caption{}
\label{fig:ladder}
\end{subfigure}

\vspace{0.5 cm}

\begin{subfigure}[b]{0.3\textwidth}
	\centering
	\begin{axopicture}(60,30)(0,0)	
	\SetWidth{1.0}
    \SetColor{Black}
	\Arc(15,15)(15,0,360)
	\Arc(45,15)(15,0,360)
	\Line(15,0)(15,30)
	\Vertex(15,0){2}
	\Vertex(15,30){2}
	\Vertex(30,15){2}
\end{axopicture}
\caption{}
\label{fig:fig8bis}
\end{subfigure}
\begin{subfigure}[b]{0.3\linewidth}
  \centering
  \begin{axopicture}(90,30)(0,0)
    \SetWidth{1.0}
   	\Vertex(30,15){2}
   	\Vertex(60,15){2}
	\Arc(15,15)(15,0,360)
	\Arc(45,15)(15,0,360)
	\Arc(75,15)(15,0,360)
\end{axopicture}
\caption{}
\label{fig:3bubble}
\end{subfigure}
\begin{subfigure}[b]{0.3\textwidth}
	\centering
	\begin{axopicture}(30,30)(0,0)	
	\SetWidth{1.0}
    \SetColor{Black}
	\Arc(15,15)(15,0,360)
	\Arc(15,0)(21.2,45,135)
	\Arc(15,30)(21.2,225,315)
	\Vertex(0,15){2}
	\Vertex(30,15){2}		
\end{axopicture}
\caption{}
\label{fig:eyeball}
\end{subfigure}

\caption{Three-loop vacuum diagrams. All diagrams can be written in terms of the basic topology (a). 
Variants of (d) and (e) with the self-loop attached in different ways are not shown.}
\label{fig:3loop}
\end{center}
\end{figure}

\begin{figure}
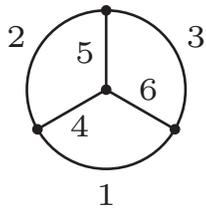

	\centering
	\scalebox{2}{
	\begin{axopicture}(30,40)(0,0)
	\tiny	
	\SetWidth{0.5}
    \SetColor{Black}
	\Arc(15,15)(15,0,360)
	\Line(15,30)(15,15)
	\Line(15,15)(2,7.5)
	\Line(15,15)(28,7.5)
	\Vertex(15,15){1}
	\Vertex(15,30){1}
	\Vertex(2,7.5){1}
	\Vertex(28,7.5){1}
	\Text(15,-5){1}	
	\Text(-2,25){2}	
	\Text(32,25){3}	
	\Text(10,8){4}	
	\Text(11,22){5}	
	\Text(23,15){6}	
\end{axopicture}
}
\vspace{0.5 cm}

\caption{Basic topology for the thre-loop vacuum diagrams.}
\label{fig:tetrawithlabels}
\end{figure}

All possible three-loop vacuum diagrams (shown in figure \ref{fig:3loop}) can be expressed in terms of a single topology (see figure \ref{fig:tetrawithlabels} for the labels of the propagators)
\be
I_{n_1...n_6}\equiv \prod_{\ell=1}^3\int\frac{d^dq_\ell}{(2\pi)^d}\prod_{i=1}^6\frac{1}{(k_i^2+u_i)^{n_i}}
\ee
where $k_{1,2,3}=q_{1,2,3}$, $k_4=q_1-q_2$, $k_5=q_2-q_3$, $k_6=q_3-q_1$.

Any three-loop vacuum graph can be written in this way (potentially with some $n_i=0$). 
\footnote{Notice that some naively different topologies (such as the diagram in figure \ref{fig:ladder}) require the use of some algebra, i.e.
\be
\frac{1}{k_1^2-u_1}\frac{1}{k_1^2-u_1'}=\frac{1}{u_1-u_1'}\left(\frac{1}{k_1^2-u_1}-\frac{1}{k_1^2-u'_1}\right)
\ee
etc. }
With all masses different, one has 47 master integrals, and consequently there are six $47 \times 47$ matrices $A_i$. These matrices  have to be computed only once, and can be used for the reduction of any three-loop vacuum diagram of arbitrary mass configuration.

Let's say we are interested in the special mass configuration $u_i=u$. By obvious symmetries, only 6 of the 47 original  master integrals are independent, one possible set is
\be
\{I_{001011},\ I_{001111},\ I_{011101},\ I_{011102},\ I_{011111},\ I_{111111}\}|_{u_i=u}
\label{eq:massesequal}
\ee
However, we have the following IBP identity between the original set of master integrals, 
\be
u_2 I_{021101}+u_3 I_{012101}+u_4 I_{011201}+u_6 I_{011102}=\frac{8-3d}{2} I_{011101}
\ee
Not all of these integrals belong to the original set of 47 master integrals, so it cannot be used to reduce this set further.
However, when all masses are equal, $u_i=u$, we have the trivial symmetry relations $I_{021101}= I_{012101}= I_{011201}= I_{011102}$ and hence
\be
4 uI_{011102}=\frac{8-3d}{2} I_{011101}
\label{eq:euler3loop}
\ee
Now, both  integrals in eq.~(\ref{eq:euler3loop}) are  in the  list of master integrals eq.~(\ref{eq:massesequal}), so we have found a new nontrivial symmetry relation between master integrals that involved the use of IBP identities.
For automatization purposes, symmetry relations such as eq.~(\ref{eq:euler3loop}) are much harder to find systematically (in contrast to the trivial symmetry relations that do not involve IBP identities). The condition in eq.~(\ref{eq:An1I0}) 
automatically detects these relations and hence can be very helpful to find them in practice.

A minimal set of master integrals can thus be taken as
\be
\vec J=\begin{pmatrix}I_{001011}\\ I_{001111}\\ I_{011101}\\ I_{011111}\\ I_{111111}\end{pmatrix}_{u_i=u}
\ee

A convenient limit direction is given by $(1,-1,0,0,0,0)$, and  
the resulting matrix $Q(t)$ is now a $47\times 5$ dimensional matrix that can easily be computed to any given order in $t$.

\section{Introducing the code \tt{MERLIN} }
\label{sec:merlin}

\subsection{Main functions}

\texttt{MERLIN} is written in \texttt{Wolfram Mathematica} and is compatible with version 12.0 or later, and is avaibalble at  \texttt{\url{https://github.com/MERLIN-QFT/MERLIN-1.0.00}}.  The current version provides pre-built connection matrices $A_i$ for evaluating generic vacuum bubble diagrams with two and three loops, one loop diagrams with up to three external momenta, and  two loop diagrams with up to two external momenta (we plan to extend this library considerably in future versions).
The program uses the known master integrals of the corresponding Feynman integral families as a basis to construct the matrices $A_i$ required for  our method. These master integral datasets were generated using \texttt{FIRE 6.5.2} \cite{Smirnov:2019qkx,Smirnov:2023yhb}.

The \texttt{MERLIN} package can be loaded with the following commands:
\begin{Verbatim}[frame=single, framesep=5mm]
SetDirectory[NotebookDirectory[]];
Get["packages/MERLIN.wl"];
\end{Verbatim}

A minimal example of usage is:
\begin{Verbatim}[frame=single, framesep=5mm]
INITIALIZE["3-loops-vacuum"]
MASSCONFIG[u, u, w, w, w, w];
DIAGRAM[{2, 1, 1, 1, 1, 1}];
EVALUATE;
\end{Verbatim}

The program requires three pre-built input files to operate correctly: the matrices, the master integrals, and the incidence matrix. These files are already provided for the two-loop and three-loop vacuum cases.

The main user-level functions are:
\begin{itemize}
\item[•] \texttt{INITIALIZE[]} — Loads the initial data: matrices, master integrals, and incidence matrix that were built for a given topology (e.g., \texttt{"3-loops-vacuum"}).
\item[•] \texttt{MASSCONFIG[]} — Specifies the mass configuration for the diagrams. It also reduces the list of 
master integrals applying the symmetries, as well as any implicit symmetries.
\item[•] \texttt{DIAGRAM[]} — Specifies the diagram to be evaluated. The function automatically detects multiple entries, given as lists.
\item[•] \texttt{EVALUATE} — Evaluates the chosen diagrams using the currently loaded data and mass configuration.
\end{itemize}

After a successful evaluation, the program saves each result in a separate \texttt{.dat} file, named using the structure \texttt{NAME\{diagram\}\{MASSCONFIG\}}, where \texttt{NAME} refers to the entry specified in the \texttt{LOAD} command. The results are then automatically loaded and stored in the output variable \texttt{RESULTS}.  
It is important to note that the \texttt{EVALUATE} function checks whether the diagram, mass configuration, and initial data belong to the same topology and will automatically load any previously saved result\footnote{To refresh all results, the corresponding data files must be deleted from the \texttt{results} folder.}.

\subsection{Internal Structure Overview}

The core evaluation routine, \texttt{EVALUATE}, performs all computational steps required to obtain the result for a given diagram.  
Internally, it constructs the  coefficients of the chosen diagram, applies explicit and implicit symmetry relations, determines a valid limit direction for the expansion, and performs the corresponding series expansions of the coefficients, master integrals, and matrices.  

The symmetry treatment in \texttt{MERLIN} is fully automated.  
Explicit symmetries, arising from the topology and mass configuration, as well as implicit symmetries, are handled through dedicated internal routines.  
This systematic reduction eliminates redundant master integrals and simplifies subsequent evaluations.

\subsection{Examples}

\subsubsection{Two-loop vacuum diagrams}
Consider the two-loop example in Sec.~\ref{sec:twoloopex}. To obtain the result, one needs to use the following commands:
\begin{Verbatim}[frame=single, framesep=5mm]
INITIALIZE["2-loops-vacuum"]
DIAGRAM[{2, 2, 1}];
MASSCONFIG[u, u, w];
EVALUATE;
\end{Verbatim}

After the evaluation is complete, the results for the chosen diagram are stored in the variable \texttt{RESULTS}:
\begin{Verbatim}[frame=single, framesep=5mm]
((2-d) G[{0,1,1}])/(2 u (4 u-w))+((-2+d) G[{1,1,0}])/(2 u (4 u-w))
-((-3+d) G[{1,1,1}])/(4 u-w)
\end{Verbatim}

Here, the functions \texttt{G[\{\}]} denote the master integrals.

\subsubsection{Three-loop vacuum diagrams}
For the three-loop example, e.g.  with all masses equal, one needs to use the following commands:
\begin{Verbatim}[frame=single, framesep=5mm]
INITIALIZE["3-loops-vacuum"]
MASSCONFIG[u,u,u,u,u,u];
\end{Verbatim}
After executing this command, one obtains the following results for both the reduced master integral list and the implicit symmetry relations:

Reduced master integral list (\texttt{MASTERRED}):
\begin{Verbatim}[frame=single, framesep=5mm]
{G[{0,0,1,0,1,1}],G[{0,0,1,1,1,1}],G[{0,1,1,1,0,1}],
G[{0,1,1,1,0,2}],G[{0,1,1,1,1,1}],G[{1,1,1,1,1,1}]}
\end{Verbatim}

Implicit symmetries (\texttt{IMPLICITRULES}):
\begin{Verbatim}[frame=single, framesep=5mm]
{G[{0,1,1,1,0,2}]-> -(((-8+3 d) G[{0,1,1,1,0,1}])/(8 u))}
\end{Verbatim}

%
%
%
%

\section{Conclusions}
\label{sec:conclusions}

In this paper we have introduced a new method to reduce a large class of Feynman integrals to linear combinations of master integrals.
The method proceeds by covariant differentiation on the space dual to the one spanned by the master integrals, with connections $A_i$ that have to be computed only once for a given topology, and is then useable for arbitrary internal mass configurations.
The limit to the mass configuration of interest is then obtained by a simple power series expansion in one auxiliary variable.

We have implemented the method in the code \texttt {MERLIN}.
Various improvements of our code are planned for future versions.
\begin{itemize}
\item
The present version contains the connections $A_i$ for simple topologies such as the two and three-loop vacuum graphs and some simple one and two-loop non-vacuum graphs. We plan to expand this library considerably for future versions. 
\item
Our main algorithm involves only simple algebraic operations (differentiation and matrix multiplication), these operations are very fast and the only time consuming step is the simplification of intermediate/final results. 
We have tried various smarter  routines (for instance, by adding more intermediate simplifications). However, there seems to be no universal choice that speeds up all cases, usually one gains in some cases and looses in others.
We however expect to be able to improve the performance considerably by an adaptive simplification algorithm that identifies the optimal choice for each case separately. This is also left for future work.
\item
Another desirable feature would be an integration with FIRE (or similar codes) in order to provide an  interface for the creation of the matrices $A_i$, for instance  if the user would like to use a different basis than the one provided or by going beyond the library of topologies included in the distribution. This will also be added in a future version.
\item
The method is not yet suitable to handle general irreducible scalar products (ISPs), besides those that appear already in the master integrals themselves. This may be resolved by adding auxiliary mass variables for the ISPs.
\end{itemize}

\section*{Acknowledgements}
GG acknowledges financial support by the Conselho Nacional de Desenvolvimento
Científico e Tecnológico (CNPq) under fellowship number 313238/2023-5, as well as 
the Fundação de Amparo à Pesquisa do Estado do Rio de Janeiro (FAPERJ) under 
project number 210.785/2024. VL is supported by Coordenação de Aperfeiçcoamento de Pessoal de Nível Superior (CAPES).

\appendix

\section{\texttt{MERLIN} Internal Structure}

The \texttt{EVALUATE} function performs all steps required to obtain the result for a given diagram. It relies on the following internal routines:
\begin{itemize}
\item[•] \texttt{PREPAREDIFFERENTIAL} — Generates the linear combination of master integrals for the chosen diagram ~\eqref{eq:X}.
\item[•] \texttt{SYMMETRY} — Automatically determines the explicit symmetries of the diagram, see Sec.~\ref{sec:covdiff}.
\item[•] \texttt{IMPLICITSYM} — Identifies implicit (non-topological) symmetries, see Sec.~\ref{sec:covdiff}.
\item[•] \texttt{FINDLIMITDIRECTION} — Determines a valid \emph{limit direction} for expansions, ~\eqref{eq:limitdir}.
\item[•] \texttt{DIFFSERIESEXPANSION} — Computes the series expansion of the coefficients generated by \texttt{PREPAREDIFFERENTIAL}, ~\eqref{eq:X-expansion}
\item[•] \texttt{SERIESEXPANSION} — Expands the master integrals and matrices, ~\eqref{eq:Iexp} and ~\eqref{eq:Aexp} .
\end{itemize}

\texttt{MERLIN} automatically handles symmetries in vacuum diagrams, via \texttt{SYMMETRY} and \texttt{IMPLICITSYM}, from these two function ones obtain
\begin{enumerate}[label=(\alph*)]
\item \emph{Explicit symmetries} — Evident from the topology and mass configuration, \texttt{SYMMETRY}.
\item \emph{Implicit symmetries} — Subtler relations not directly apparent from topology, \texttt{IMPLICITSYM}
\end{enumerate}
Their results can be inspected through:
\begin{enumerate}[label=(\alph*)]
\item \texttt{MASTERRED} — The reduced list of master integrals with symmetries applied.
\item \texttt{IMPLICITRULES} — The set of implicit symmetry rules used in the final result.
\end{enumerate}
\texttt{MERLIN} determines a suitable limit direction automatically using \texttt{FINDLIMITDIRECTION}.  
This function generates a vector of the same length as \texttt{MASSCONFIG} and tests whether the mass limit \texttt{MASSCONFIG + t * LIMITDIRECTION} introduces singularities or indeterminate expressions.  
If no singularities are found, the vector is stored as the \texttt{LIMITDIRECTION} and used in the series expansion.  
Note that this procedure depends only on the mass configuration and the matrices, not on the specific diagram.

\bibliographystyle{JHEP}

\bibliography{references}

\end{document}